\documentclass[conference]{IEEEtran}
\usepackage{amsmath,amsfonts}
\usepackage[utf8]{inputenc}
\usepackage{multirow}
\usepackage[dvipsnames]{xcolor}

\usepackage{algorithm}
\usepackage[noend]{algpseudocode}
\usepackage{array}
\usepackage[caption=false,font=normalsize,labelfont=sf,textfont=sf]{subfig}
\usepackage{textcomp}
\usepackage{stfloats}
\usepackage{url}
\usepackage{verbatim}
\usepackage{graphicx}
\usepackage{cite}
\usepackage{tabularx}
\newcolumntype{C}{>{\centering\arraybackslash}X}
\usepackage{listings}
\algrenewcommand\algorithmicindent{0.5em}%

\newcommand{\toolname}{SAT-MapIt}

\newcommand\blfootnote[1]{%
  \begingroup
  \renewcommand\thefootnote{}\footnote{#1}%
  \addtocounter{footnote}{-1}%
  \endgroup
}

\begin{document}

\title{\toolname:\\A SAT-based 
Modulo Scheduling Mapper \\for Coarse Grain Reconfigurable Architectures 
}

\author{
\IEEEauthorblockN{Cristian Tirelli}
\IEEEauthorblockA{\textit{SYS Institute} \\
\textit{Università della Svizzera italiana}\\
Lugano, Switzerland \\
cristian.tirelli@usi.ch}
\and
\IEEEauthorblockN{Lorenzo Ferretti}
\IEEEauthorblockA{\textit{Computer Science} \\
\textit{University of California, Los Angeles}\\
Los Angeles, United States \\
ferrelo@cs.ucla.edu
}
\and
\IEEEauthorblockN{Laura Pozzi}
\IEEEauthorblockA{\textit{SYS Institute} \\
\textit{Università della Svizzera italiana}\\
Lugano, Switzerland \\
laura.pozzi@usi.ch}
}

\maketitle

\begin{abstract}
Coarse-Grain Reconfigurable Arrays (CGRAs) are emerging low-power architectures aimed at accelerating compute-intensive application loops.
The acceleration that a CGRA can ultimately provide, however, heavily depends on the quality of the mapping, i.e. on how effectively the loop is compiled onto the given platform. State of the Art compilation techniques achieve mapping through modulo scheduling, a strategy which  attempts to minimize the II (Iteration Interval) needed to execute a loop, and they do so usually through well known graph algorithms, such as Max-Clique Enumeration.

We address the mapping problem through a SAT formulation, instead, and thus explore the solution space more effectively than current SoA tools.
To formulate the SAT problem, we introduce an ad-hoc schedule called the \textit{kernel mobility schedule} (KMS), which we use in conjunction with  the data-flow graph and the architectural information of the CGRA in order to create a set of boolean statements that describe all constraints to be obeyed by the mapping for a given II. We then let  the SAT solver efficiently navigate this complex space. As in other SoA techniques, the process is iterative: if a valid mapping does not exist for the given II, the II is increased and a new KMS and set of constraints is generated and solved.

Our experimental results show that \toolname\ obtains better results compared to SoA alternatives in $47.72\%$ of the benchmarks explored: sometimes finding a lower II, and others even finding a valid mapping when none could previously be found.

\end{abstract}


\section{Introduction}
\blfootnote{This work was supported by the Swiss National Science Foundation under Grants 200020-182009 and  200020-188613, and in part by the National Science Foundation under Grants P2TIP2\_199735.}The constant growth of computational requirements in everyday applications has increased the demand for high-performance and low-power architectures, able to perform compute-intensive tasks efficiently while at the same time dealing with tight power/resource constraints.

While Application Specific Integrated Circuits (ASICs) accelerators have been largely adopted in these scenarios due to their efficiency, they are limited by fixed functionality.
due to their fine-grained structure.

\begin{figure}[t]
\centering
  \includegraphics[width=0.25\textwidth]{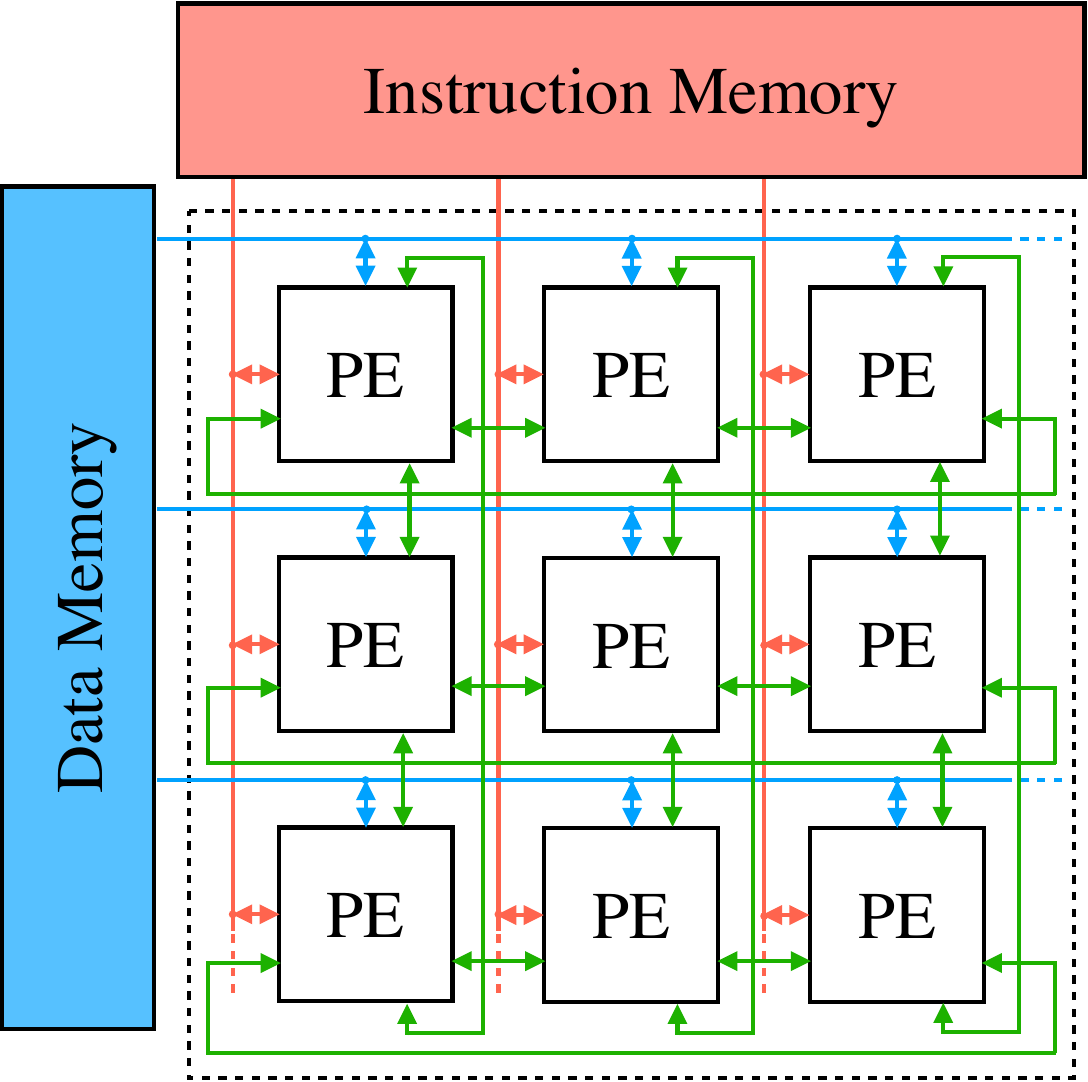}
  \caption{Example $3\times 3$ CGRA. Processing elements are connected in a 2D mesh with near-neighbour topology. 
  }
  \label{fig:cgra}\vskip -2em
\end{figure}

Coarse-Grain Reconfigurable Arrays (CGRAs),
which are programmable architectures able to achieve high efficiency with low power requirements\cite{li2021chordmap}\cite{karunaratne2017hycube}, provide a meet-in-the-middle approach given their coarse-grain reconfigurability, and hence have become popular in various domains such as streaming and multimedia applications \cite{li2021chordmap, akbari2018px, karunaratne2017hycube, oh2009recurrence, lee2015optimizing, wijerathne2019cascade}, 
 as well as medical ones
\cite{duch2017heal}. 
A CGRA is a mesh of Processing Elements (PE) organized in a two-dimensional grid; each PE contains an ALU (Arithmetic-Logic Unit) and a number of internal registers, and is connected to neighbors PEs according to its mesh topology, and to main memory through memory lines. 
Figure \ref{fig:cgra} shows the generic CGRA architecture targeted by our methodology.

One of the fundamental challenges of CGRA exploitation is the compilation process, i.e. the translation of high level source code onto CGRA code, while taking advantage of the parallelism offered by the architecture. To do so, a technique called modulo scheduling is traditionally employed, which constructs a DFG from an application loop body, and then maps DFG nodes onto architecture PEs in an interleaved manner so that nodes from different iterations coexist in the same cycle, minimising the overall latency of each iteration.

Existing techniques for modulo scheduling, described in Section \ref{sec:related_works}, rely mainly on heuristics to schedule, place and then route instructions and data on the PEs. 
Herein, we propose to address the mapping problem using SAT-MapIt, a SAT-based formulation where data dependency, architectural constraints, and schedule are expressed as Boolean constraints. A valid mapping is then identified by determining if an assignment of the variables exists to satisfy the constructed boolean formula.
We show that this technique is able to efficiently explore the space of possible mappings better than SoA techniques, hence producing high-performance schedules.


\section{Related Work}\label{sec:related_works}

Existing methodologies for the CGRA mapping problem can be divided in two main categories: the first using heuristics, the second using exact solutions. 
Early works in the first category include the method proposed by Mei et al.\cite{mei2007adres}, which formulates scheduling, placement, and routing problems altogether, and proposes to solve it using simulated annealing. 
Alternatively, in \cite{park2008edge} an edge-centric approach to modulo-scheduling was presented, where a schedule is generated by first routing each edge in the dataflow graph, and placement is addressed as a by-product of routing. In \cite{hamzeh2012epimap}, EPImap proposed to use both routing and re-computation to find a valid mapping, by adopting an epimorphic (time-extended) graph formulation.


Epimap's performance was later improved by GraphMinor \cite{chen2014graph} and REGIMap \cite{hamzeh2013regimap},  by reducing the mapping problem to the graph minor and max clique problem. In turn, \cite{dave2018ramp} RAMP authors have further refined REGIMap by  explicitly modeling and exploring various routing strategies and choosing the best one for each given loop kernel. CRIMSON\cite{balasubramanian2020crimson} then proposed a randomized iterative modulo scheduling algorithm that explored the scheduling space more efficiently, and PathSeeker \cite{pathseeker} improved on~\cite{balasubramanian2020crimson} by analyzing mapping failures and performing local adjustments to the schedule to obtain a lower compilation time and a better quality of the solution.

In our experiments we compare our results to those obtained by both RAMP\cite{dave2018ramp} and PathSeeker\cite{pathseeker}. These two works had shown superior performance with respect to the earlier methodologies mentioned above, and therefore represent the current SoA of the modulo scheduling mapping problem. We quantitatively compare our work to them, and show that \toolname\ can better explore the scheduling space and get smaller II (Iteration Interval) by using custom scheduling tables with a SAT formulation.

A second category of approaches addresses the mapping problem  with Integer Linear Programming (ILP) or Boolean Satisfiability formulations. In \cite{chin2018architecture} the authors propose an ILP formulation approach and prove the feasibility of mapping in the given number of cycles. Similarly, \cite{miyasaka2020sat} propose to use a SAT solver instead of an ILP solver to identify a valid solution. 
The work proposed in \cite{miyasaka2020sat} represents a first effort towards exact formulations using a SAT solver; however, it is not capable of achieving a modulo scheduled solution.
Our SAT formulation is, to the best of our knowledge, the first to propose an exact solution to the Modulo Scheduling problem, and yet scale to Data Flow Graph sizes that were previously only tackled via heuristics.

\section{Background}\label{sec:background}

In this section we provide the background needed to present our methodology, and we illustrate it wherever appropriate through a running example.
\subsection{Compilation}

In order to accelerate an application onto a CGRA, a compute-intensive loop is identified in the application; the identification can either be performed automatically via techniques such as for example \cite{zacharopoulos2018regionseeker}, or manually by the programmer via pragma-annotations, as done in this work. Then, the loop needs to be compiled, in order to be translated onto CGRA instructions.



The first step in this process is to generate a  semantically-equivalent version in Intermediate Representation (LLVM IR in our case) and from there to Data Flow Graph (DFG) depicted in Figure \ref{fig:pke}a. DFGs are directed graphs in which nodes represent instructions, edges represent dependency relations between instructions, and back-edges correspond to loop-carried dependencies.


In a second phase, the generated DFG is mapped onto the CGRA, by assigning each of its nodes to a given PE at a given cycle. In order to perform such mapping, a technique called modulo scheduling is employed.

%

\subsection{Modulo Scheduling}


\begin{figure}[t]
    \centering
    \includegraphics[width= 0.18\textwidth]{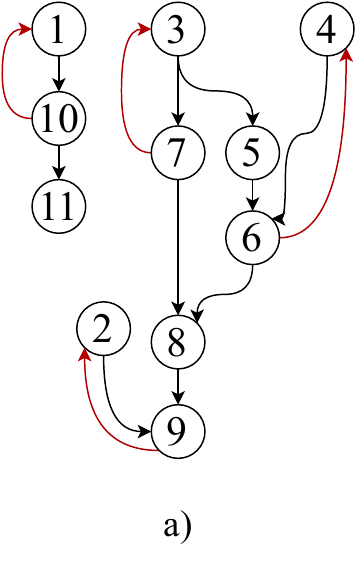}
    \includegraphics[width= 0.265\textwidth]{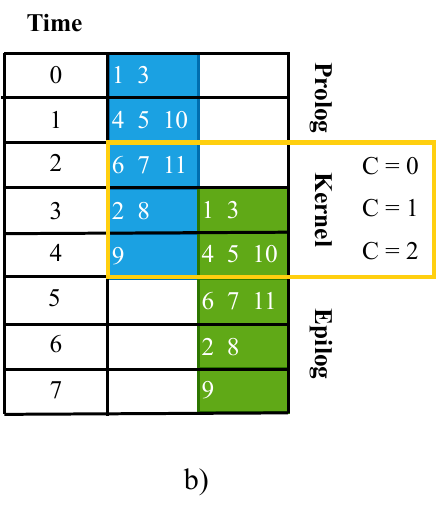}
    \includegraphics[width= 0.36\textwidth]{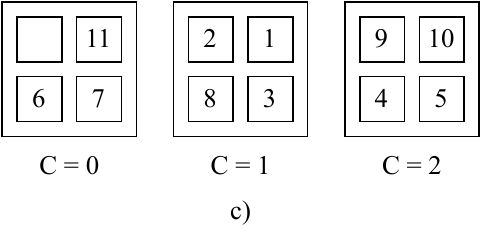}
    \caption{a) Example Data Flow graph of a loop. b) Modulo scheduling of the DFG on the left, highlighting the division among Prolog, Kernel, and Epilog. The loop is unrolled once; the first iteration is in blue, and the second,  shifted by $II$, is in green. c) Mapping example on a $2\times 2$ CGRA}
    \label{fig:pke}\vskip -1em
\end{figure}

Modulo Scheduling (MS) is a compilation technique that enables efficient execution of a loop body, by executing multiple iterations of it in an interleaved manner. As exemplified in Figure \ref{fig:pke}b, a modulo-scheduled loop is divided into three stages: prologue, kernel, and epilogue. Prologue and epilogue are one-time executed stages: the former is used to prepare the data to feed the pipeline, the latter to reorganize them at the end. The kernel is instead repeated multiple times and includes the instructions to be parallelized through pipelining.
Goal of MS is to  pipeline as effectively as possible the execution of kernel instructions, and this corresponds to minimizing the Iteration Interval (II), i.e. the length of the kernel stage, which is 3 cycles in our example.
The mapping problem hence consists in finding a legal Modulo Schedule for a loop,  performing the placement and routing of instructions in a constrained 3D space represented by the PE dimensions and by time. 
An example of legal mapping for the DFG in the running example is shown in Figure \ref{fig:pke}c.





\section{Methodology}\label{sec:methodology}

\subsection{Overview}\label{sec:overview}
\begin{figure}[t]
\centering
\includegraphics[width=0.45\textwidth]{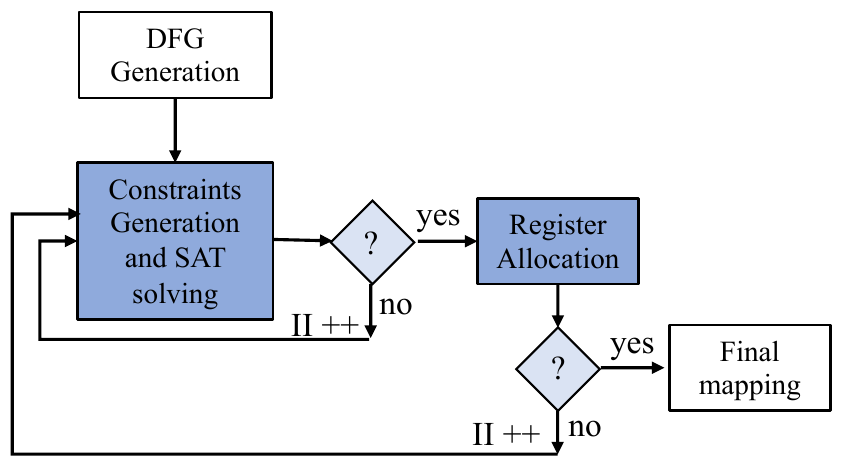}
\caption{\toolname\ searches for mappings for a given $II$, iteratively increasing $II$ in case the SAT solver returns UNSAT, or register allocation fails to colour the  model returned by the solver.}
 \label{fig:toolchain}\vskip -2em
\end{figure}

Our methodology addresses mapping by solving a SAT problem where data dependency, schedule and CGRA architecture are expressed as Boolean constraints in a conjunctive normal form (CNF).
Our toolchain, depicted in Figure \ref{fig:toolchain}, takes as input the C code of the application, converts it into LLVM IR, extracts the designated loop structures and, through a custom LLVM pass, generates its DFG.


The information retrieved from the LLVM pass
is then used to build a set of schedules, in turn translated into a set of constraints, which get finally fed to a SAT solver.
 If the answer of the solver is SAT, the next step is to verify that there are enough registers available in the PEs to store the generated data for the whole liveness duration, and this is determined via Register Allocation. If this step succeeds, then a valid mapping has been identified. Otherwise, the current Initiation Interval is increased and the process is iteratively repeated.



Herein, we detail the various steps of our methodology: schedule creation, SAT formulation, and register allocation.

\subsection{Schedule creation}\label{subsec:schedule_creation}

We first create As-Soon-As-Possible (ASAP) and As-Late-As-Possible (ALAP) schedules, for the input DFG. This corresponds to detecting how early and how late each node can be scheduled. 
We then generate the Mobility Schedule (MS), which expresses the mobility of each node from its $ASAP$ to its $ALAP$ time position.
Figure \ref{fig:scheds} shows these schedules for our running example. 

\begin{figure}[h]
\centering
\small
\begin{tabular}{|c|lll|}
\hline
              & \multicolumn{1}{c||}{\textbf{ASAP}} & \multicolumn{1}{c||}{\textbf{ALAP}} & \multicolumn{1}{c|}{\textbf{MS}} \\ \hline
\textbf{Time} & \multicolumn{3}{c|}{\textbf{Nodes}}                                                                        \\ \hline
0             & \multicolumn{1}{l||}{1 2 3 4}       & \multicolumn{1}{l||}{3}             & 1 2 3 4                          \\ \hline
1             & \multicolumn{1}{l||}{5 7 10}        & \multicolumn{1}{l||}{4 5}           & 1 2 4 5 7 10                     \\ \hline
2             & \multicolumn{1}{l||}{6 11}          & \multicolumn{1}{l||}{1 6 7}         & 1 2 6 7 10 11                    \\ \hline
3             & \multicolumn{1}{l||}{8}             & \multicolumn{1}{l||}{2 8 10}        & 2 8 10 11                        \\ \hline
4             & \multicolumn{1}{l||}{9}             & \multicolumn{1}{l||}{9 11}          & 9 11                             \\ \hline
\end{tabular}
\vspace{0.2cm}
\vskip -1em
\caption{ASAP, ALAP, and Mobility Schedule}
\label{fig:scheds}
\end{figure}

\begin{figure}[h]
    \centering
    \includegraphics[width= 0.48\textwidth]{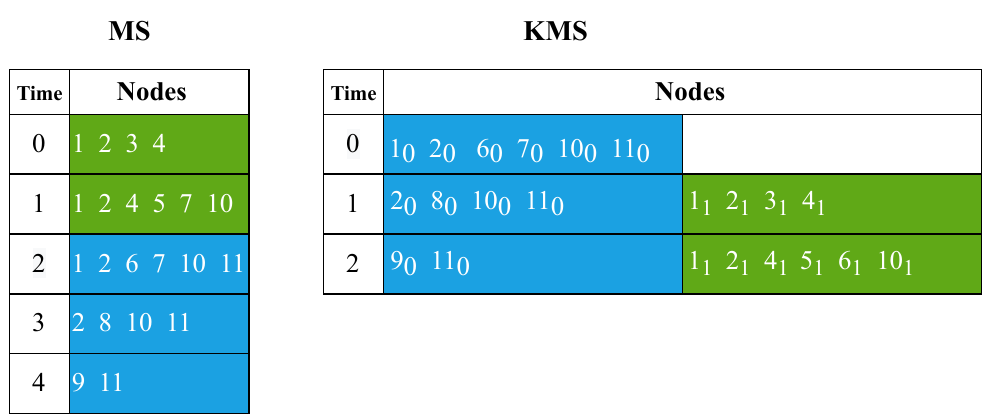}
    \caption{Kernel Mobility Schedule creation. \\ In blue iteration 0, in green iteration 1}
    \label{fig:kmsgen}\vskip -1em
\end{figure}

At this point we create the Kernel Mobility Schedule (KMS): a custom structure which \toolname\ uses to then formulate the mapping problem.
The KMS  can be seen as \emph{a superset of all possible kernels}, and is the product of iteratively folding MS by an amount equal to $II$: every time MS is folded by $II$ into KMS, each node receives a label that refers to the iteration number it belongs to. The more folding we have, the more iterations our kernel will execute at the steady-state. This process is depicted in Figure \ref{fig:kmsgen}, again for our running example. Given an $II$ of 3, the MS is folded twice ($\lceil 5/3 \rceil = 2$) and hence the KMS contains two iterations. The first is depicted in blue, and the second in green.

Together, DFG and KMS are used to generate all the statements of the CNF formulation of the mapping problem, which is the subject of the next subsection. 
\vskip -1em


\subsection{SAT formulation}

We create a CNF formula using literals in the form:
$x_{n, p, c, it}$ , where $n$ denotes the node identifier in the DFG, $p$ denotes a PE on the CGRA, $c$ represents at which cycle a node is scheduled, and $it$ to  which iteration the node refers to. 

Our problem formulation can be described at a high level by partitioning all statements into three main sets of clauses that assure the following:
\begin{itemize}
\item C1: Every node is associated with a set of literals, and for each one of those sets, one and only one literal must be set to \textit{True}.
\item C2: At most one node should be assigned to a PE at a given cycle, since two or more nodes cannot be scheduled simultaneously on the same PE.
\item C3: Each node's predecessor and/or successor must be assigned to a neighbor or on the same PE.
\end{itemize}
To provide a formal description of the above constraints, we introduce additional definitions.
Let $\mathcal{L}$ be the set of all literals, then $\mathcal{L}(n)$ is the set of all literals associated to node $n$. 

To make the notation more compact and easy to read, we also associate each literal
in the form $x_{n,p,c,it}$ to a literal written as $v_i$.
For example $\mathcal{L}(n_3)$ would be written as:
\begin{equation*}
    \mathcal{L}(n_3)= \{v_{0},v_1,v_2,v_3\}
\end{equation*}
where $v_0 = x_{3,0,1,1}$, $v_1 = x_{3,1,1,1}$, $v_2 = x_{3,2,1,1}$ and $v_3 = x_{3,3,1,1}$.
This represents the fact that node 3 appears only at time 1 and only at iteration 1 in the KMS, as shown in Figure \ref{fig:kmsgen}, and that it can be mapped onto any PE: 0,1,2,3.
Now we can start the description of the three sets of constraints. The first set, C1, ensures that all nodes are mapped on the CGRA, and can be encoded formally with:
\begin{equation}
	\begin{aligned}
		\phi(n) &= \bigvee_{v_i \in \mathcal{L}(n)} v_i \\
		\xi(n) &=\bigwedge_{(v_i,v_j)\in \mathcal{M}(n)} \neg (v_i \wedge v_j)\\
		\zeta(n) &= \phi(n) \wedge \xi(n)\\ 
\label{eq:stm1}
	\end{aligned}
\end{equation}
\vskip -1em
where $n$ is one of the nodes id in the DFG and $\mathcal{M}(n)$ is defined as follows:
\begin{equation*}
    \mathcal{M}(n) = \{(v_i,v_j): v_i \prec v_j, (v_i,v_j)\in \mathcal{L}(n) \times \mathcal{L}(n) \}
\end{equation*}

where $v_i = x_{n,p_1,c_1,it_1}$ and $v_j = x_{n,p_2,c_2,it_2}$ with 
$p_1 \neq p_2$, $c_1\neq c_2$ and $it_1\neq it_2$.
Furthermore the symbol $\prec$ represents the \textit{lexicographically smaller-than} relation between two literals.
For example $x_{3,0,1,0}$ is lexicographically smaller than $x_{3,1,0,0}$, while $x_{3,1,0,1}$ is not.

Equation \ref{eq:stm1} will be used on each node of the DFG and each $\zeta$ generated will be added to the SAT formulation.

The second set of constraints, C2, forbids the mapping of more than a single node  on the same PE at the same time, and is encoded through:
\begin{equation}
	\begin{aligned}
	    M(n,m) &= \bigwedge_{(v_i,w_j)\in \mathcal{V}(n,m)} \neg (v_i \wedge w_j) \\
		\zeta &= \bigwedge_{n}^{N - 1} \bigwedge_{m = n + 1}^{N} M(n,m)\\
	\end{aligned}
	\label{eq:diffpe}
\end{equation}
with $\mathcal{V}(n,m)$ defined as:
\begin{equation*}
    \mathcal{V}(n,m) = \{(v_i,w_j): v_i \prec w_j, v_i\in \mathcal{L}(n), w_j \in \mathcal{L}(m)\}
\end{equation*}

The last set of constraints, C3, handles the dependencies in the DFG, and assures that 1) each data dependency is mapped on neighbors PE on the CGRA and that 2) data produced by a node in a given cycle is consumed before being overwritten by another node on a subsequent cycle.
For each dependency, we consider  literals that are at most one iteration apart in the KMS, that are on a neighbour PE and that respect one of the relations:
\begin{equation}
    c_d \leq c_s \text{ if } it_s \neq it_d \quad or \quad
    c_d > c_s \text{ if } it_s = it_d
    \label{eq:cc}
\end{equation}
where $c_d$ is the cycle at which the destination node is scheduled, and $c_s$ is the cycle at which the source node is scheduled. 
This constraint ensures that a node consumes the value produced by the predecessor in the proper order, avoiding overlapping of the same dependencies among kernel and prologue/epilogue stages.

The CNF in this case is composed of two main terms; one that handles the case in which the output of the source node is delivered to the destination node through the registers on the PE, and the other in which the data created by the source node is written in the output register of the PE, instead, and hence it must not be overwritten in subsequent cycles.

Dependencies that use internal registers can be encoded through a set of CNF of this form:
\begin{equation}
    \zeta_1 = v_i \wedge w_j
    \label{eq:c31}
\end{equation}
where $v_i = x_{n_s, p_1,c_1, it_1}$ and $w_j = x_{n_d, p_2,c_2, it_2}$ and $p_1$ is neighbour of $p_2$.
On the other hand dependencies that do not use internal registers need to be encoded with a slightly different formulation. In particular we expand Equation \ref{eq:c31} by appending another term.
The CNF in this case become:
\begin{equation}
    \zeta_2 = v_i \wedge w_j \wedge \neg \left( \bigvee z_k\right)
    \label{eq:c32}
\end{equation}
where $z_k = x_{n_l, p1, c_l, it_l}$ iterate over all literals in all subsequent cycles, with the same PE as source node. This assures that the data is properly consumed by the destination node without being overwritten.
Equation \ref{eq:c31} and \ref{eq:c32} are repeated for each dependency in the KMS and every CNF generated is added to the formulation 
after an \textit{or} operation among all the terms, so that the solver is free to choose the option that makes the mapping problem satisfiable.



\subsection{Register allocation}\label{subsec:register_allocation}


Once the solver returns a satisfying mapping, a last phase is needed in order to validate such mapping in terms of register usage: Register Allocation --  implemented in \toolname\ as a separate phase subsequent to SAT solving.
It is solved optimally by exploiting the SSA format of the input code\cite{hack2006register}. For each PE in the CGRA, we generate an interference graph that we proceed to color. If the coloring succeeds, no further action is needed, and the coloring hence generated concludes the mapping, by adding information on which register should hold the output of each PE at each cycle.
If it fails, however, 
we split the overlapping intervals that make the interference graph uncolorable with the given number of colors, by adding load and stores, and hence additional cycles.

\begin{figure*}[htp]
\centering
\includegraphics[width=\textwidth]{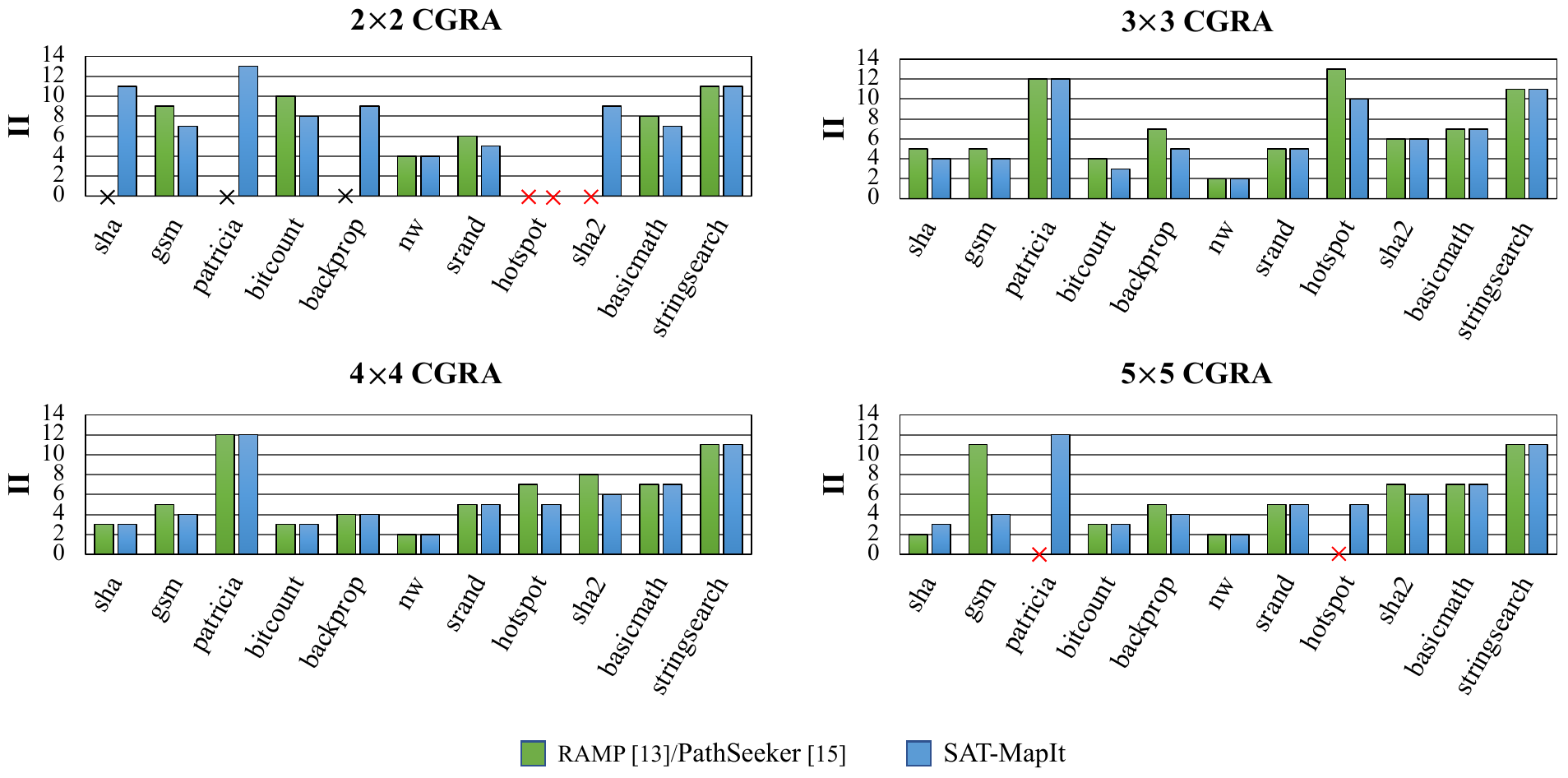}\hfill

\caption{Experimental results of the chosen benchmarks for different architecture sizes.
We compare the II found by \toolname\ with respect to the best results obtained by RAMP and PathSeeker -- lower is better. A red mark means that the process did not terminate before a timeout of 4000 seconds. 
A black mark means that the process was terminated when it reached a current II of 50 but still found no feasible solution.
}
\label{fig:benchs}\vskip -1em

\end{figure*}
\begin{table*}
    \begin{minipage}{\columnwidth}
    \centering
    \begin{tabular}{|l|l|l|l|l|}
\hline
\textbf{Benchmarks} & \textbf{{[}13{]}/{[}15{]}} & \textbf{SAT-MapIt}  & \multicolumn{1}{c|}{$\Delta$} \\ \hline
\texttt{sha}                 & 41.01                                      & 3.22               & -37.79              \\ \hline
\texttt{gsm}                 & 2.81                                       & 1.25               & -1.56              \\ \hline
\texttt{patricia}            & 1351.15                                    & 5.39               & -1345.76              \\ \hline
\texttt{bitcount}            & 2.63                                       & 1.68               & -0.95              \\ \hline
\texttt{backprop}             & 1262.69                                    & 3.39               & -1259.3              \\ \hline
\texttt{nw}                  & 0.01                                       & 0.56               & 0.55              \\ \hline
\texttt{srand}               & 0.32                                       & 1.15               & 0.83              \\ \hline
\texttt{hotspot}            & 4000                                        & 4000               & 0              \\ \hline
\texttt{sha2}                & 4000                                       & 2.21               & -3997.79              \\ \hline
\texttt{basicmath}           & 0.01                                       & 0.62               & 0.61              \\ \hline
\texttt{stringsearch}        & 0.19                                       & 1.02               & 0.83              \\ \hline
\end{tabular}
\caption{Mapping time (seconds) on a $2\times 2$ CGRA}
    \label{tab:2x2}
    \end{minipage}\hfill 
    \begin{minipage}{\columnwidth}
    \centering
    \begin{tabular}{|l|l|l|l|l|}
\hline
\textbf{Benchmarks} & \textbf{{[}13{]}/{[}15{]}} & \textbf{SAT-MapIt}  & \multicolumn{1}{c|}{$\Delta$} \\ \hline
\texttt{sha}                 & 0.23                                       & 2.86              & 2.63                \\ \hline
\texttt{gsm}                 & 4.14                                       & 4.35              & 0.21                \\ \hline
\texttt{patricia}            & 48.98                                      & 16.31              & -32.67                \\ \hline
\texttt{bitcount}            & 5.86                                       & 7.84              & 1.98                \\ \hline
\texttt{backprop}             & 44.62                                      & 12.27              & -32.35                \\ \hline
\texttt{nw}                  & 0.03                                       & 1.56              & 1.53                \\ \hline
\texttt{srand}               & 0.09                                       & 3.9              & 3.81                \\ \hline
\texttt{hotspot}             & 13.19                                      & 28.43              & 15.24                \\ \hline
\texttt{sha2}                & 61.7                                       & 3.13              & -58.57                \\ \hline
\texttt{basicmath}           & 0.07                                       & 1.9              & 1.83                \\ \hline
\texttt{stringsearch}        & 3.27                                       & 3.55              & 0.28                \\ \hline
\end{tabular}
\caption{Mapping time (seconds) on a $3\times 3$ CGRA }
    \label{tab:3x3}
    \end{minipage}
    \begin{minipage}{\columnwidth}
    \centering
    \begin{tabular}{|l|l|l|l|l|}
\hline
\textbf{Benchmarks} & \textbf{{[}13{]}/{[}15{]}} & \textbf{SAT-MapIt}  & \multicolumn{1}{c|}{$\Delta$} \\ \hline
\texttt{sha}                 & 32.04                                      & 7.23             & -24.81               \\ \hline
\texttt{gsm}                 & 32.04                                      & 10.46             & -21.58               \\ \hline
\texttt{patricia}            & 421.05                                     & 39.28             & -381.77               \\ \hline
\texttt{bitcount}            & 0.44                                       & 21.55             & 21.11               \\ \hline
\texttt{backprop}             & 57.17                                      & 25.1             & -32.07               \\ \hline
\texttt{nw}                  & 0.08                                       & 3.63             & 3.55               \\ \hline
\texttt{srand}               & 0.25                                       & 8.79             & 8.54               \\ \hline
\texttt{hotspot}             & 3556.62                                    & 3734.77             & 178.15               \\ \hline
\texttt{sha2}                & 696.6                                      & 7.19             & -689.41               \\ \hline
\texttt{basicmath}           & 0.22                                       & 4.11             & 3.89               \\ \hline
\texttt{stringsearch}        & 0.02                                       & 7.62             & 7.6               \\ \hline
\end{tabular}
\caption{Mapping time (seconds) on a $4\times 4$ CGRA}
    \label{tab:4x4}
    \end{minipage}\hfill 
    \begin{minipage}{\columnwidth}
    \centering
    \begin{tabular}{|l|l|l|l|l|}
\hline
\textbf{Benchmarks} & \textbf{{[}13{]}/{[}15{]}} & \textbf{SAT-MapIt}  & \multicolumn{1}{c|}{$\Delta$} \\ \hline
\texttt{sha}                 & 6.25                                       & 28.93             & 22.68              \\ \hline
\texttt{gsm}                 & 0.29                                       & 21.4             & 21.11              \\ \hline
\texttt{patricia}            & 4000                                       & 75.16             & -3924.84             \\ \hline
\texttt{bitcount}            & 0.01                                       & 47.62             & 47.61              \\ \hline
\texttt{backprop}             & 981.73                                     & 52.43             & -929.3              \\ \hline
\texttt{nw}                  & 0.02                                       & 7.75             & 7.73              \\ \hline
\texttt{srand}               & 0.02                                       & 21.64             & 21.62              \\ \hline
\texttt{hotspot}             & 4000                                    & 108.02             & -3891.98              \\ \hline
\texttt{sha2}                & 675.12                                     & 16.88             & -658.24              \\ \hline
\texttt{basicmath}           & 0.5                                        & 8.7             & 8.2              \\ \hline
\texttt{stringsearch}        & 0.02                                       & 15.3             & 15.28              \\ \hline
\end{tabular}
\caption{Mapping time (seconds) on a $5\times 5$ CGRA}
    \label{tab:5x5}
    \end{minipage}
\end{table*}

\section{Experimental Results}
\label{sec:results}

\textbf{Experimental Setup.} We evaluate the effectiveness of \toolname\ on a set of loop kernels from MiBench and Rodinia benchmark suites. 
We compare the obtained $II$, and the time to find it,  with respect to two techniques of the SoA: RAMP\cite{dave2018ramp} and PathSeeker \cite{pathseeker}. For these two SoA techniques we use the original code publicly released by the authors.
In the target CGRA architecture that we consider in our experiments, each PE is connected to the four nearest neighbors, as in Figure \ref{fig:cgra}, and each PE contains four local registers. We vary the size of the mesh from $2\times 2$ up to $5\times 5$.
The Z3 solver 
is used to solve our SAT formulation. All experiments are performed on a machine with 2.6 GHz 6-Core Intel Core i7. For PathSeeker, each experiment was repeated 10 times given its randomized nature.


\textbf{\toolname\ achieves better IIs}. The performance of a mapping is first and foremost measured by the $II$ achieved, because this, in turn, is a measure of the level of parallelism obtained. In our experiments we compare the $II$ of \toolname\ with those of SoA, for each benchmark explored. This is depicted in Figure \ref{fig:benchs} which shows the performance obtained by all techniques for different CGRA configurations. For SoA, we report the best result among the two algorithms. Our tool is able to systematically find the best mapping solution. 


\textbf{\toolname\ uses tight resources better.} By focusing on the $2\times 2$  size CGRA, we can see that SAT-MapIt always finds the best solution, and twice (\texttt{patricia} and \texttt{backprop}) it is even able to find a valid mapping where the SoA could not. This showcases the effectiveness of our methodology particularly when the mapping problem becomes more challenging.



\textbf{\toolname~is faster when runtimes are high.}
Given that the $II$ found are better than SoA, we now analyse the time needed to find such solutions, and report it in Tables  \ref{tab:2x2}, \ref{tab:3x3},  \ref{tab:4x4} and  \ref{tab:5x5}. 
It can be noticed that our tool running time is longer than SoA in 26 out of 44 experiments, and that in these 26 cases the average time difference is only $15.28$ seconds, with a standard deviation of $34.97$. On the other hand, in the 18 cases in which our tool is faster, the average time difference is $962.24$ seconds, with a standard deviation of $1438.78$. This shows that \toolname\ is significantly faster when it matters, i.e. when computation times are high. 

\textbf{Limitations of \toolname}
Currently, our tool does not apply any routing strategy. This limitation manifests in the \texttt{sha} kernel of a $5\times 5$ CGRA, where we achieve an $II$ of 3, while SoA can find an $II$ of 2 by adding a routing node. This is the only case, out of the 44 experiments shown, where the effect of this limitation can be noticed. 




\section{Conclusion}\label{sec:conclusion}
In this paper we present a tool, called \toolname, for modulo-scheduling loops onto CGRAs. We find that previous techniques, mainly based on classic graph algorithms such as Max-Clique enumeration, do not always explore the scheduling space effectively. We propose a new SAT formulation of the modulo scheduling problem on CGRA that fully explores the scheduling space and finds the lowest II possible for a given DFG.
To define the mapping problem through a SAT formulation, we introduce a new custom schedule called Kernel Mobility Schedule, which is used with the data-flow graph of the loop to be mapped, and with the architectural information of the CGRA, to generate all the constraints that the SAT solver needs to obey. 
Overall, \toolname\ finds better solutions than the SoA alternatives\cite{dave2018ramp}\cite{pathseeker}, achieving better results in $47.72\%$ of the cases, and even identifying valid mappings where other tools could not find a valid solution. 

\bibliographystyle{IEEEtran}
\bibliography{biblio} 

\end{document}